\documentclass[preprint,12pt]{elsarticle}
\usepackage{lineno}
\usepackage{multirow}
\usepackage{xcolor}
\usepackage{float}
\usepackage[utf8]{inputenc}
\usepackage[T1]{fontenc}
\usepackage{mathtools}
\usepackage[thinc]{esdiff}
\usepackage[bottom]{footmisc}
\usepackage{float}
\usepackage{amsmath}
\usepackage[euler]{textgreek}
\usepackage{siunitx}
\usepackage{graphicx}% Include figure files
\usepackage{dcolumn}% Align table columns on decimal point
\usepackage{bm}% bold math
\usepackage{array}
\usepackage{caption}
\usepackage{subcaption}
\usepackage{tabularx}
\usepackage{soul}
\usepackage{xcolor, soul}
\sethlcolor{yellow}
\renewcommand{\arraystretch}{1.8}

\usepackage{amssymb}
\begin{document}

\begin{frontmatter}



\title{Uncertainties in the Transport Properties of Helium Gas at Cryogenic Temperatures Determined Using Molecular Dynamics Simulation}


\author[inst]{Kaanapuli Ramkumar}

\affiliation[inst]{organization={Department of Physics, Indian Institute of Technology Kanpur},%Department and Organization
            city={Kanpur},
            postcode={208016}, 
            state={Uttar Pradesh},
            country={India}}

 \author[inst]{Swati Swagatika Mishra}

\author[inst]{Sudeep Bhattacharjee}
%\author[inst1,inst1]{Author Three}

\begin{abstract}
In this study, the transport properties, such as diffusivity, viscosity, and thermal conductivity, of \textsuperscript{4}He at the gaseous phase are computed for state points in the temperature range of 10 K to 150 K and pressure range of 0.10 MPa (1 atm) to 0.21 MPa using classical and quantum frameworks. Within the classical molecular dynamics simulation (MDS), the Green-Kubo (GK) approach is used. The GK method has an inherent uncertainty associated with it due to the random fluctuations in the flux autocorrelation functions. Moreover, in the temperature range of 10 K to 40 K, the quantum nature of the helium gas particles becomes prominent. The classical MDS performed does not include these effects and hence introduces uncertainties in the calculated results. We discuss efficient ways of time averaging the autocorrelation function to reduce statistical fluctuations and perform the quantum scattering phase shift calculations within the Chapman-Cowling theory to study the quantum effects. Furthermore, this study also provides the transport properties data in the cryogenic temperature limit of 10 K to 150 K, which is scarcely studied in the literature and can be applied in complex systems.
\end{abstract}




\end{frontmatter}
%\linenumbers

\section{Introduction}
Knowledge of transport properties of simple gases to complex fluidic systems is important in various interdisciplinary fields such as fuel combustion technologies, nuclear reactor engineering, extraterrestrial explorations and nuclear fusion including plasmas technologies. Transport coefficients such as diffusion, thermal conductivity, and shear viscosity that quantify the transfer of energy, momentum, and mass in a fluid, act as a bridge connecting the microscopic molecules in a fluid with the macroscopic system as a whole. The interaction and the properties of the individual molecules in a fluid determine the transport coefficients, which can be measured at the macroscopic level. The design of thermal protection systems for planetary exploration probes, modelling of atmospheres of other planets, designing inertial confinement fusion reactors and the modelling of plasma processes used in the gasification of biomass requires extensive data on the transport properties of various species. Furthermore, the transport properties of neutral gases become very important for the understanding of the dynamics of charged species in a weakly ionized plasma medium, especially at cryogenic temperatures (4 K to 10 K) and at atmospheric pressure \cite{cryointro,cryoplasma}.

In recent years, transport properties of different systems have been studied in detail employing molecular dynamics simulations (MDS). The advantage of using simulation is that it has flexibility and allows one to manipulate any parameter of a system, thereby permitting one to understand the effects of different parameters in greater detail without the limitations of conventional experimental investigations \cite{econo1,econo2,vrabec,vrabec2,EVANS19821}. Takahashi \textit{et al.} studied the quantum effect of liquid hydrogen on bubble nucleation \cite{mds_quantH}. Yiannourakou \textit{et al.} performed an MDS study to optimize the design and performance of acid gas removal units \cite{mds_acid}. González-Salgado \textit{et al.} studied the anomalous behaviour of water at 4\textsuperscript oC with the help of MDS \cite{mds_temp}. Cunha and Robbins studied the pressure-viscosity relation of 2,2,4-trimethylhexane using MDS \cite{mds_vis}. Recently, the temperature guided behavioural transitions in confined helium at cryogenic temperatures (30 K to 150 K) were looked at by our group \cite{MISHRA2022100073}. In the current work, we extend the investigation to even lower temperatures where the quantum effects would be important.

All these studies are a testimony to the importance of MDS in understanding physical phenomena. The most prominently used approach to calculating transport properties using MDS is the Green Kubo (GK) approach, an equilibrium molecular dynamics method, which uses the autocorrelation function of the flux, corresponding to the physical property of interest, to compute the transport properties. Non-equilibrium molecular dynamics (NEMD) methods can also be used to study the transport properties, but they have to be calculated in different directions separately, unlike the equilibrium methods \cite{NEMD1,NEMD2}. For a uniform gaseous system like the one presented in this study, global values of the transport coefficients are more desirable as all the transport coefficients have similar values in all three directions (X, Y, and Z). Therefore, we have chosen the equilibrium GK method. However, the GK approach is known to have a significant statistical error associated with it \cite{green_error,acf_exp}. Furthermore, at low temperatures and high pressure regimes, the gas molecules tend to show their quantum nature and the classical MDS are unable to provide accurate results, leading to the divergence in the classically calculated transport coefficients from the experimental values. Therefore, the quantum nature of the molecules is also a source of error while using the classical MDS \cite{Nagashima_2014,mds2_qm}. 

For the efficient usage of any simulation result in practical applications, one has to be aware of the different uncertainties involved with the simulation. The purpose of the present article is to understand the uncertainties associated with the calculation of transport properties in various temperature and pressure regimes. The primary focus of the article is on the uncertainties associated with the GK approach and the manifestation of quantum effects at low temperature regimes. A simulation performed can also suffer from systematic errors due to the choice of force field \cite{uncer3}, errors in the simulation parameters, simulation software used \cite{uncer1}, and so on. Choice of the force field should be made carefully based on the attractive and repulsive nature of the system under study \cite{MolMod}, and systematic errors are usually state point independent, giving an overall error in the simulation, much the same way as the systematic errors in experiments. This makes the reproducibility of simulation data by different groups difficult even when the same simulation parameters are used \cite{uncer2,uncer4}. As opposed to the systematic errors, we focus on the uncertainty that arises from a physical process, such as the emergence of the quantum nature of gas particles at low temperatures, and the uncertainty innate to the widely used GK approach due to the random fluctuations in the autocorrelation functions of the thermodynamic fluxes. We also address the issue of the effects of long-range interactions that arise from the attractive tail of the force field.

For this purpose, we use the Lennard-Jones Truncated (LJT) potential, which is extensively used in literature owing to its simple mathematical form parameterised by two variables and its ability to accurately describe the force field of noble gases \cite{EVANS19821,uncer4,ljpot1,ljpot2,lr2}. We study the transport coefficients such as viscosity ($\eta$), thermal conductivity ($\kappa$), and diffusivity ($\mathcal{D}$) of \textsuperscript 4He in the gas phase within a temperature range of 10 K to 150 K and pressure range of 0.10 MPa (1 atm) to 0.21 MPa, avoiding supercritical phases. These low temperature state points, at atmospheric pressures, have not been extensively studied to the best of our knowledge. The limited number of molecular dynamics studies at this temperature and pressure regime is due to the large uncertainty that arises while using the GK method for low density gases. The GK method used is best suited for dense fluids, such as liquids, as the decay of the autocorrelation function is slower with a decrease in the density of the system studied. This increases the computation time of the MDS and also the error due to fluctuations in the autocorrelation function. We performed the simulation several hundred times, details are given in Appendix A, and the transport coefficient datas are given in Tables \ref{tab:V_data}, \ref{tab:K_data} and \ref{tab:D_data}.

\begin{figure*}[t]
\centering
\includegraphics[trim= 0 0 0 0, clip,scale=0.2]{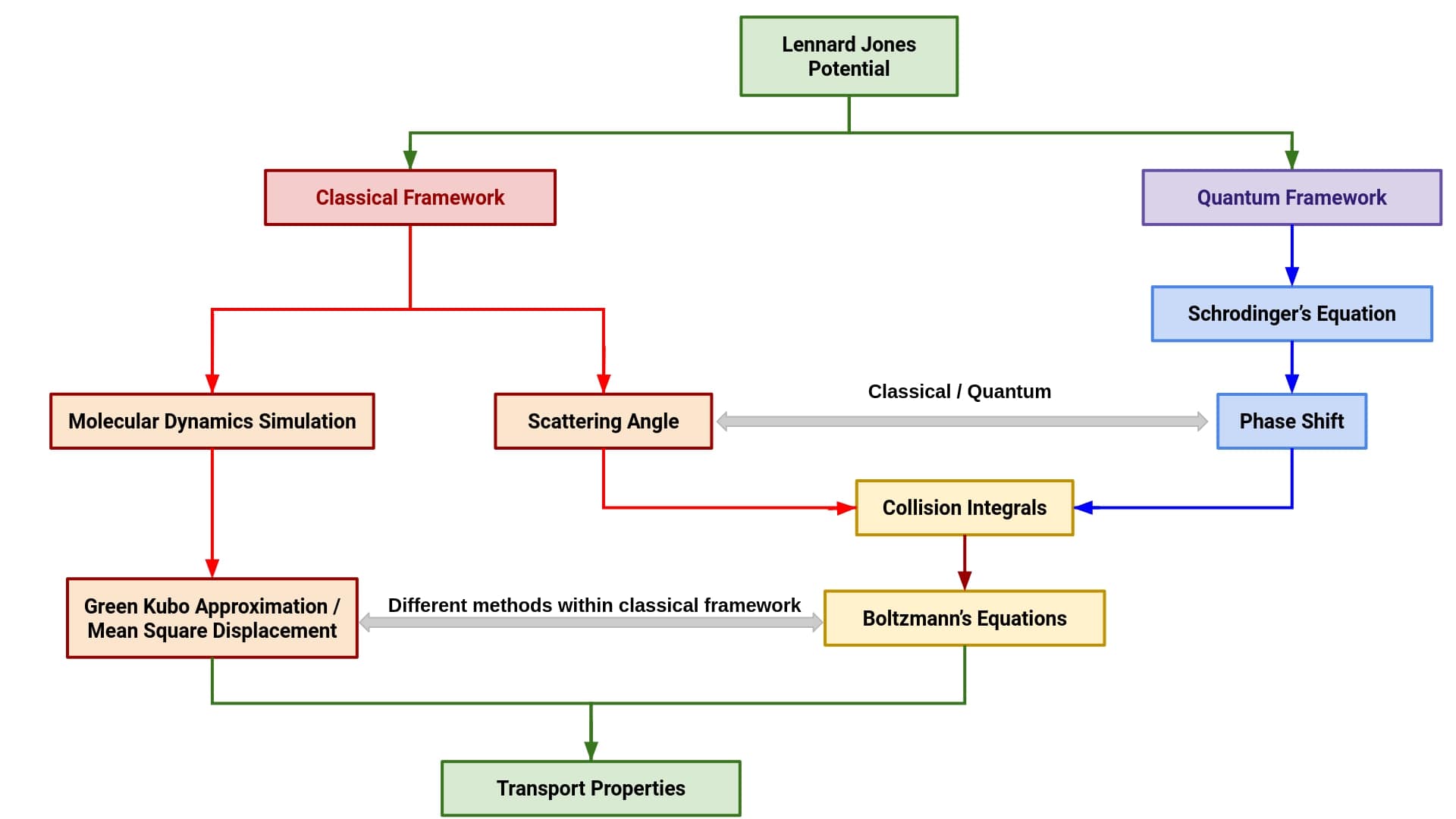}\label{fig:flow}
\caption{Calculation of transport coefficients within different frameworks. }
\end{figure*}

This article is divided into five sections. Section 2 briefly describes the theoretical concepts of calculating transport coefficients and quantum effects manifesting at low temperatures. Section 3 contains the details of the MDS simulations and quantum phase shift calculations. Section 4 discusses the results of the simulation and quantum calculations, and finally, the article is concluded in Section 5.

\section{Theoretical Background}

\subsection{Transport Properties}

The article studies transport properties of \textsuperscript 4He gas in a temperature regime where the quantum effects start to emerge (around 40 K). Hence, it is necessary to study them in both the classical and the quantum formalisms, as the investigated temperature range in this work is from 10 K to 150 K. Within the classical formalism, the transport properties could be calculated in two ways: (i) by simulation methodologies such as MDS, using the Green-Kubo method \cite{green,kubo} and (ii) through Boltzmann transport equations employing  Chapman-Cowling approximations \cite{Ferziger1972MathematicalTO}. For simple systems such as noble gases where binary interactions are predominant, one could use Boltzmann transport equations, which involve the calculation of scattering angles in binary collisions to be used in Chapman-Enskog theory and Chapman-Cowling approximations \cite{Ferziger1972MathematicalTO}. In quantum formalism, we use the same Chapman-Enskog theory and Chapman-Cowling approximations used in the classical method, but the Schr\"odinger's equation is solved to calculate the phase shifts of scattering processes, which plays the quantum analogous role of scattering angle computed in the classical formalism. Henceforth, we call the Chapman-Cowling approximations used within the quantum framework as the quantum phase shift calculation method. The flow chart in Fig. \ref{fig:flow} depicts how these different methodologies are employed in the calculation of the transport coefficients, and the following subsections briefly summarise them.

\subsubsection{Green Kubo Theory}
The GK relations \cite{green,kubo},  use the autocorrelation functions (ACF) of the flux corresponding to the transport properties to evaluate the transport coefficients. The viscosity ($\eta$) is calculated from the ACF of the off-diagonal pressure tensor ($P_{ij}(t)$), given by \cite{viscosity},
\begin{equation}\label{eq:viscosity}
    \eta = \frac{V}{k_B T}\int_0^{\infty} \langle P_{ij}(t_0)P_{ij}(t+t_0)\rangle_{t_0} dt,
\end{equation}
the thermal conductivity ($\kappa$) from the heat flux ($J(t)$) ACF, given by \cite{thermalcon},
\begin{equation}\label{eq:thermal_conductivity}
    \kappa = \frac{V}{3k_B T}\int_0^{\infty} \langle J(t_0)\cdot J(t+t_0)\rangle_{t_0} dt,
\end{equation}
and the diffusion coefficient ($\mathcal D$) from the velocity ($v(t)$) ACF, given by \cite{allen},
\begin{equation}\label{eq:diffusion}
    \mathcal{D} = \frac{1}{3}\int_0^{\infty} \langle v(t_0)\cdot v(t+t_0)\rangle_{t_0} dt,
\end{equation}
where $t_0$ and $t$ are any arbitrary initial and instantaneous times, respectively.

\subsubsection{Chapman-Cowling Approximation}

To obtain the transport properties from the Chapman-Cowling approximation, the scattering angles for binary collisions between the gas molecules are used to evaluate the collision cross-sections given by  \cite{Ferziger1972MathematicalTO}
\begin{equation}\label{eq:sctt_cross}
    Q^{(r)}=2\pi\int_0^\infty\left(1-cos^r\zeta\right)\ b\ db,
\end{equation}
where $\zeta$ is the scattering angle and $b$ is the scattering parameter and $r$ takes the values 1 and 2. The collision cross-section thus evaluated is used to calculate the collision integrals ($\Omega$), given by \cite{Ferziger1972MathematicalTO},
\begin{equation}\label{eq:collision_integral}
    \Omega^{(r,s)} = \left(\frac{k_BT}{\pi m}\right)^{1/2}\int_0^\infty Q^{(r)}e^{-g^2}g^{2s+3}dg,
\end{equation}
where $s$ takes the values 1 and 2 and $g$ is the non-dimensional relative velocity of the scattering particles, given by $g=\hbar\ k/\sqrt{2k_BT\mu}$, where $\mu$ is the reduced mass and $k$ is the wave number of the particles. The Boltzmann's transport equations are solved using the Chapman-Enskog theory \cite{chapman1958mathematical}, and the transport coefficients are obtained using the Chapman-Cowling approximation, to the first order, as \cite{chapman1958mathematical},
\begin{subequations}\label{eq:chap_eq}
\begin{gather}
    \eta=\frac{5k_BT}{8 \Omega^{(2,2)}},\\
    \kappa=\frac{5k_BT}{8\Omega^{(2,2)}}\frac{15k_B}{4m},\\
    \mathcal D=\frac{3(k_BT)^2}{8P m\Omega^{(1,1)}}.
\end{gather}
\end{subequations}
Here $m$ is the mass of the interacting particles, and $P$ corresponds to the pressure of the system, in SI units, at which the transport coefficients are calculated. 

In quantum mechanics, due to the uncertainty principle, the relation between the scattering angle ($\chi$) and the impact parameter ($b$) cannot be defined to calculate the collision cross-sections (Eq. \ref{eq:sctt_cross}). Classically, when a particle scatters off of a target, it is characterised by the deflection angle, i.e., the scattering angle. In the quantum framework, due to the wave nature of the particle, the scattering of a particle induces a phase shift in the particle's wave function. Hence, to do a quantum mechanical calculation, instead of scattering angles, the phase shifts ($\delta_l$) in the asymptotic region of scattering state wave functions are used to obtain the quantum mechanical collision cross-sections. For details of scattering cross-sections in quantum mechanics, one can refer to \cite{sakurai,merzbacher1998quantum}. The quantum mechanical collision cross-sections are given by \cite{Ferziger1972MathematicalTO,DEBOER1955381,bird},
\begin{equation}\label{eq:sc_1}
    Q^{(1)}=\frac{4\pi}{k^2}\sum_{l=0}^{\infty}(l+1)\:sin^2(\delta_{l+1}-\delta_l),
\end{equation}
and
\begin{equation}\label{eq:sc_2}
    Q^{(2)}=\frac{2\pi}{k^2}\sum_{l=0}^{\infty}\frac{(l+1)(l+2)}{l+3/2}\:sin^2(\delta_{l+2}-\delta_l),
\end{equation}
which can be used in the collision integral equation (Eq. \ref{eq:collision_integral}) to evaluate the transport coefficients from the Chapman-Cowling expressions (Eqs. \ref{eq:chap_eq}). Here, $k$ is the wave number, and $l$ is the angular momentum quantum number of the state considered. To include the effects of quantum statistics, the summations in the above formulas are carried out only over the even $l$ values for the Bose-Einstein case and over the odd $l$ values for the Fermi-Dirac case due to symmetry and in the case of collision between indistinguishable particles, the formulas are multiplied by a factor of 2 \cite{Ferziger1972MathematicalTO,DEBOER1955381}. A detailed description of the quantum effects is provided in the next subsection.

\subsection{Quantum Effects}
As the principles of quantum mechanics and classical physics are inherently different, and the latter being only an approximation to the former, studying physical phenomena in the classical framework leads to missing out on some details of the process under study. In this section, the quantum effects that are important for the transport properties are briefly described.

Due to the wave-particle duality of quantum particles, at low temperatures, when the thermal de Broglie wavelength ($\lambda_{th}$), given by the expression \newline $\lambda_{th}=\sqrt{2\pi\hbar^2/(k_BmT)}$, where $k_B$ is the Boltzmann's constant, $m$ is the mass of the molecule, $T$ is the temperature, and $\hbar$ is the reduced Planck's constant, is comparable to the particle's dimensions, diffraction effects become dominant \cite{bird}. Classically, as particles are not associated with any wave nature, these diffraction effects (inherent to waves) are absent. Furthermore, at cryogenic temperatures, the particles with spin half follow Fermi-Dirac statistics, and particles with integer spin follow Bose-Einstein statistics and not the classical Boltzmann distribution. When $\lambda_{th}$ is comparable to the inter-particle distances, the effects due to the quantum statistics (symmetry effects) become dominant \cite{bird}. 

\begin{figure}
\centering
\includegraphics[trim= 20 0 40 30, clip,scale=0.2]{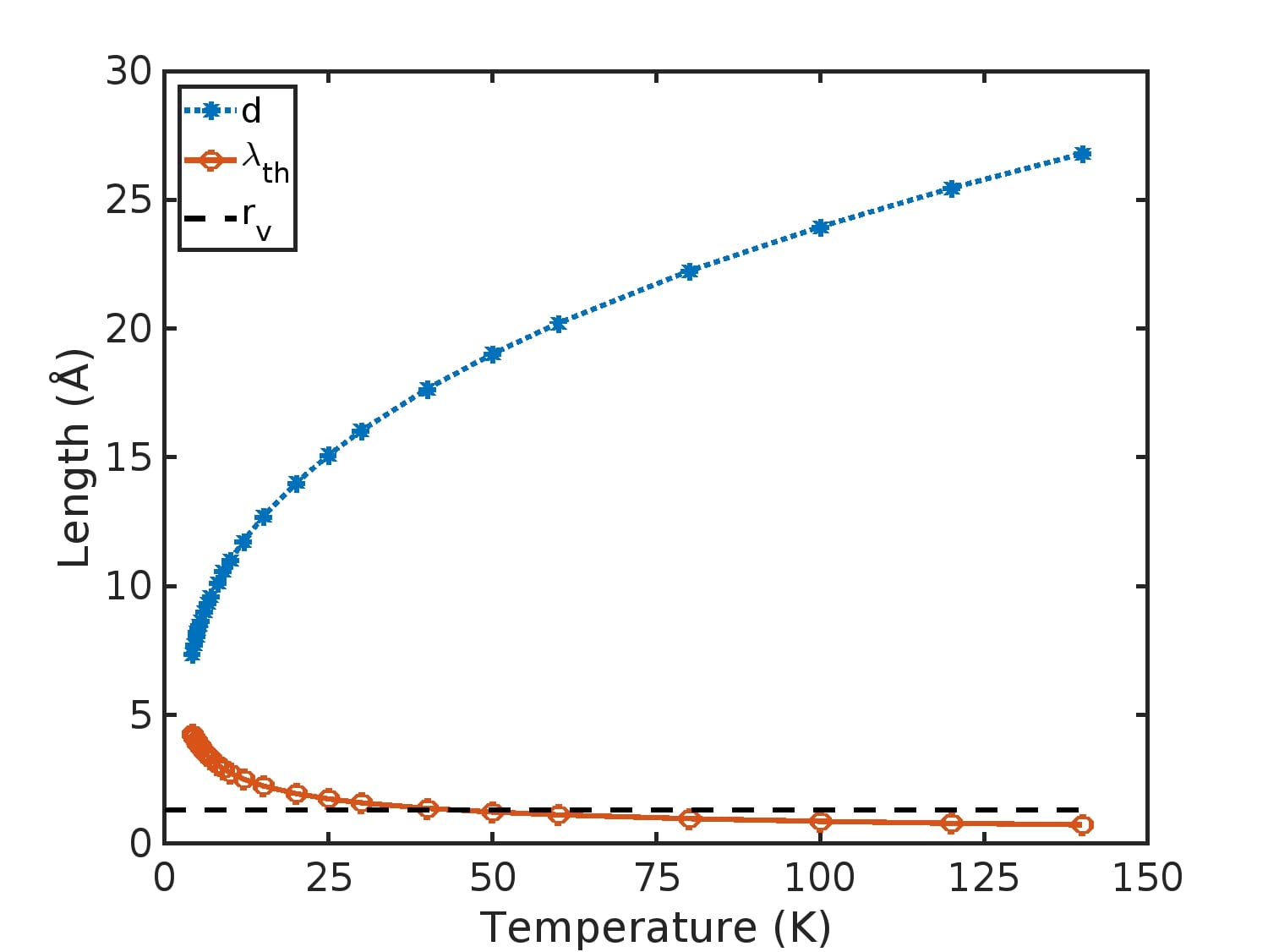}\label{fig:length}
\caption{Comparing Thermal de Broglie wavelength ($\lambda_{th}$) (solid line) with mean inter-particle distance ($d$) (dotted line). The Van der Waals radius ($r_v$) is marked by the dashed line for reference. }
\end{figure}

For \textsuperscript 4He atoms, the Van der Waals radius ($r_v$) is 1.4 \AA. This corresponds to $\lambda_{th}$ of \textsuperscript 4He at $\sim$ 45 K. Fig. \ref{fig:length} shows how the mean inter-particle distances ($d$) in \textsuperscript 4He gas, obtained from the number density ($n$) data at atmospheric pressure \cite{he_data}, from the relation $ d=n^{-1/3}$, approaches $\lambda_{th}$ at a much lower temperature, less than 4.222 K, which is the boiling point of \textsuperscript 4He gas at atmospheric pressure \cite{he_data}. Hence, while considering quantum effects, quantum statistical effects may be neglected for the temperature range of 10 - 150 K.

It should be noted that under the quantum framework, the calculation of the diffusion coefficient needs special considerations. Classically, one can track the motion of individual particles to find the self-diffusion coefficient but quantum mechanically, when identical particles interact, they lose their identity after the interaction (indistinguishability) \cite{chapman1958mathematical,bird,spin}. This causes the quantum mechanically obtained values for self-diffusion scattering cross-section to be almost twice that of classically obtained cross-section and surprisingly, the classical values are closer to the experimental values than the quantum mechanical values, even at the cryogenic temperatures. This is because experimentally it becomes impossible to track the motion of individual identical particles and hence, some property of the particles, which do not affect the collision process, is used to differentiate the particles in the system \cite{self_diff}. For instance, in the case of \textsuperscript 3He self-diffusion experiments, the molecules are categorised based on their spin state (+½ and -½) \cite{spin} as spin does not affect collision interactions. Similarly, different isotopes with mass ratios very close to unity are also used in calculating self-diffusion coefficients experimentally \cite{self_diff_iso}. By doing this, the interacting particles are made distinguishable, and the experimentally measured self-diffusion coefficients are not that of true self-diffusion coefficients of identical particles but that of distinguishable particles. To compensate for the absence of indistinguishability in the experimental procedures, theoretically, one can calculate these values by assuming a distinguishable scattering process where the self-diffusion coefficient is calculated as a limiting case of binary diffusion \cite{bird}.
\section{Computational Details}

\subsection{Interaction Potential}

In all the calculations done in this article, Lennard-Jones Truncated (LJT) potential is considered to depict the interaction between the \textsuperscript 4He atoms \cite{ljpot1,ljpot2}, which is given by
\begin{equation}\label{eq:potential}
    V_{LJ}(r) = \begin{cases}
        4\epsilon\left[\left(\frac{r}{\sigma}\right)^{-12}-\left(\frac{r}{\sigma}\right)^{-6}\right],&r\leq r_c\\
        0,&r>r_c
    \end{cases}
\end{equation}
with the energy parameter $\epsilon = \num{9.44}$ meV, the length parameter $\sigma=\num{2.64}$ $\si{\angstrom}$ \cite{ab} and cutoff length $r_c = 10\,\si{\angstrom}$. 

\subsection{Classical Molecular Dynamics Simulation}

The Large-scale Atomic/Molecular Massively Parallel Simulator (LAMMPS) \cite{LAMMPS} is used to perform the MDS. While the GK relations (Eq. \ref{eq:viscosity},\ref{eq:thermal_conductivity},\ref{eq:diffusion}) are used to calculate the transport properties,  the diffusion coefficients are also calculated from the asymptotic time slope of mean square displacement (MSD) of the particles in the system, using the Einstein's relation given by, $ \mathcal{D} = \lim_{t \to \infty}\frac{1}{6t} MSD$ \cite{allen}. This has been done to show that the uncertainties due to quantum effects are innate to classical formalism and do not depend on the exact method used to calculate the transport coefficient values. 

\subsubsection{Simulation Parameters}
The simulation system size was set at $660\si{\angstrom}\times 660\si{\angstrom}\times 660\si{\angstrom}$ and is kept constant throughout all the simulations performed in this work. The LJ potential used is truncated at a cut-off distance of 10 $\si{\angstrom}$ for all simulations, and this value is based on the prescribed cutoff length for LJ interactions, which is $r_c>2.5\sigma$ \cite{ljpot}. Beyond this cut-off length, the interaction potential is small enough (< 0.02 times $\epsilon$) and can be neglected. The time interval between the two simulation steps is 1 fs (step size) and the initial system equilibration is performed for 5 ns as an NPT ensemble and the thermalization of the system is performed under NVE ensemble setting for 1 ns. The periodic boundary conditions have been applied to the simulation system. 

Several simulations were performed to calculate the transport properties at various temperatures in the range of 10 K to 150 K for 0.10 MPa pressure (1 atm) and various other pressures in the range of 0.14 MPa to 0.21 MPa for temperatures 30 K and 100 K. Table \ref{tab:state} show the number of atoms simulated in each simulation.

\subsubsection{Long-Range Interaction Correction}

As the LJT potential is used for the computation of the transport properties, the effects of the potential beyond its cut-off length ($r_c$) are not included in the simulation. Though the long-range effects are significant in studying certain properties of the system such as surface tension and interface thickness of different phases \cite{lr_eff}, they can be neglected when studying transport properties such as viscosity, thermal conductivity, and diffusion \cite{lr2,ljpot,lr1,lr3}. However, long-range interactions lead to a correction in the pressure values, which is given by \cite{allen},
\begin{align}
    P_{crc} &= P_{sim} + P_{LRC},\notag\\
    P_{LRC} &= \epsilon\left(\frac{32}{9}\pi\rho^2\sigma^12 r_c^{-9} - \frac{16}{3}\pi\rho^2 \sigma^6r_c^{-3}\right)
\end{align}
where $\sigma$ and $\epsilon$ are the LJ parameters, $\rho$ is the number density (number of atoms per unit volume) $P_{crc}$ is the corrected pressure including the long-range correction, $P_{sim}$ is the uncorrected pressure from the simulation and $P_{LRC}$ is the long-range correction to the pressure. Table \ref{tab:state} gives the long-range correction to the pressure of the state points considered in this article. It can be seen that the long-range correction to the pressure values is two orders of magnitude lower than the pressure values considered and hence is negligible.

\subsubsection{Green Kubo ACF Integration Details}\label{sec:green_err}
The primary source of uncertainty in the GK approach is the time integral over the ACF in Eqs. \ref{eq:viscosity},\ref{eq:thermal_conductivity} and \ref{eq:diffusion} \cite{green_error}. Considering the ACF for longer time periods above the correlation length leads to increased error due to the accumulation of fluctuations in the tail of the ACF. Prematurely truncating the ACF before it sufficiently decays also leads to errors in the calculated transport properties. Various methods based on exponential fitting or appropriately truncating the ACF have been proposed to minimise this integration error \cite{viscosity,exp_fit}. In this article, to minimise the error due to random fluctuations in the ACF, the ACF is calculated as a cumulative average of a few hundred simulation runs \cite{replica1,replica2}, details of which are given in Appendix A. Also, for simple fluids in the gaseous state, the decay in the ACF is exponential in nature \cite{acf_exp}. Hence, the ACF is fit to an exponential function, $a\ e^{-t/\tau}$, where $\tau$ gives the correlation length. This removes the dependence of the transport properties on the ACF integration cutoff and any residual white noise, after the cumulative averaging, on the ACF tail.

\renewcommand{\arraystretch}{1.5} % Default value: 1
\begin{table}[H]
\caption{\centering Details of the state points considered in the simulations.}
\centering
\begin{tabularx}{1\textwidth} {|
>{\centering\arraybackslash}X   | >{\centering\arraybackslash}X   | >{\centering\arraybackslash}X   ||  >{\centering\arraybackslash}X   | >{\centering\arraybackslash}X   ||}
\hline
\multirow{2}{6em}{\centering Temperature (K)} & \multirow{2}{6em}{\centering Number of atoms (N)}& \multirow{2}{6em}{\centering Density ($\rho=N/V$) ($10^{-4}\si{\angstrom}^{-3}$)} & \multicolumn{2}{c||}{Pressure (MPa) }\\
\cline{4-5}
&&& $P_{sim}$  & $P_{LRC}\, (10^{-2})$\\
\hline
10&211090&7.34&\multirow{7}{2em}{0.10} &-1.60 \\
30 &70364&2.45&&-0.18 \\
40 &52773&1.84&&-0.10  \\ 
50 &42218&1.47&&-0.06 \\ 
70 &30156&1.05&&-0.03 \\ 
100 &21109&0.73&&-0.02 \\ 
150 &14073&0.50&&-0.01\\ 
\hline
\multirow{4}{1em}{30} 
&97221&3.38&0.14&-0.34\\
&111110&3.86&0.16&-0.44\\
&124998&4.34& 0.18&-0.56\\ 
&145831&5.07& 0.21&-0.76\\ 
\hline
\multirow{4}{1.5em}{100} 
&29166&1.01&0.14&-0.03\\
&33333&1.16& 0.16 &-0.04\\
&37499&1.30& 0.18 &-0.05\\ 
&43749&1.52& 0.21 &-0.07\\ 
\hline
\end{tabularx}
\label{tab:state}
\end{table}

\subsection{Quantum Mechanical Calculation}

One of the disadvantages of the purely classical MDS is that at low temperatures, the QM effects are not taken into account. In the case of \textsuperscript 4He gas, the QM effects start to dominate at relatively higher temperatures (at around 40 K) than for any other gases. Using the quantum mechanical theory, the transport properties can be obtained by solving for the phase shifts in the asymptotic region of the scattering state wave function, which then can be used to calculate the collision cross-section in the Chapman-Cowling relations \cite{chapman1958mathematical}. 

To obtain the phase shifts, the Schr\"odinger's equation, given by
\begin{equation}\label{eq:schrodinger}
    \frac{\partial^2 \mathcal{U}(r)}{\partial r^2} = f(r)\ \mathcal{U}(r),
\end{equation}
is solved numerically using the Numerov's integration method \cite{numerov}, where the wave function is discretised as,

\begin{equation}
    \mathcal{U}_{i+1} = \frac{\mathcal{U}_{i-1}\ (12-d^2f_{i-1})-2\ \mathcal{U}_i\ (5d^2f_i+12)}{d^2f_{i+1}-12}.
\end{equation}
Here, $d$ is the step size considered for discretizing the wave function, $l$ is the angular momentum state considered, $\mu$ is the reduced mass of two colliding \textsuperscript{4}He atoms and for the case of LJ potential, the function $f(r)$ is given by,
\begin{equation}
        f(r)=\frac{2\mu}{\hbar^2}\left( l(l+1)\frac{\hbar^2}{2\mu r^2} + V_{LJ}(r)-E\right),
\end{equation}
where $E$ is the scattering state energy eigenvalue. In our calculations, the wave functions are discretized with a step size of 0.025 a.u. and are evaluated up to a distance of 1500 a.u.

The analytical expression for the asymptotic scattering state wave function, for a given angular momentum state $l$, is given by \cite{sakurai},
\begin{equation}\label{eq:Asymptotic}
    \mathcal{U}(r)=A\:kr\ (j_l(kr)\ \cos\ \delta_l - n_l(kr)\ \sin\ \delta_l),
\end{equation}
where $A$ is a normalization factor, $\delta_{l}$ is the phase shift corresponding to a given angular momentum state $l$, $j_l$ and $n_l$ are the spherical Bessel and Neumann functions, and $\mathcal{U}(r) = R(r)\ r$, with $R(r)$ being the radial part of the scattering wave function. The phase shift ($\delta_l)$ can be evaluated using,
\begin{equation}
    \Gamma = \frac{r_2\ \mathcal{U}(r_1)}{r_1 \mathcal{U}(r_2)} = \frac{j_l(kr_1)\ \cos\ \delta_l - n_l(kr_1)\ \sin\ \delta_l}{j_l(kr_2)\ \cos\ \delta_l - n_l(kr_2)\ \sin\ \delta_l} ,
\end{equation}

and
\begin{equation} \label{eq:phase_shift}
    \delta_l =\frac{j_l(kr_1)-\Gamma\ j_l(kr_2)}{n_l(kr_1)-\Gamma\ n_l(kr_2)}.
\end{equation}

To obtain the quantum scattering cross-section, the phase shift values must be calculated for all even angular momentum states at different energies from 0 to $\infty$, as the quantum cross-sections have summation up to infinite angular momentum state (Eqs. \ref{eq:sc_1} and \ref{eq:sc_2}) and the collision integral has integral over wave number from zero to infinity (Eq. \ref{eq:collision_integral}). This is approximated by carrying out the summation over angular momentum states in Eqs. \ref{eq:sc_1} and \ref{eq:sc_2}, and integration over the wave vector, in Eq. \ref{eq:collision_integral}, until the change in the cross-section and collision integral is not more than a factor of $10^{-8}$ after the inclusion of the phase shifts for higher angular momentum states and energy values. The integration in collision integral (Eq. \ref{eq:collision_integral}) is performed using the trapezoidal method as it varies smoothly with wave number. For the temperature region of interest in this article (10 K to 150 K), a maximum angular momentum state of $l=22$ and a maximum wave vector of 9 a.u. is considered. Here, it is assumed that only binary interactions take place during a collision, i.e. only two particle collisions are considered \cite{Ferziger1972MathematicalTO,chapman1958mathematical}.
%%%%%%%%%%
\section{Results}
The values of transport coefficients, namely viscosity ($\eta$), thermal conductivity ($\kappa$), and diffusivity ($\mathcal{D}$) calculated using different methods in the temperature range of 10 K to 150 K and pressure range of 0.10 MPa to 0.21 MPa are provided in Tables \ref{tab:V_data}, \ref{tab:K_data} and \ref{tab:D_data}. The statistical error associated with the GK approach in calculating viscosity and thermal conductivity is mentioned along with their values. The diffusion coefficient, which was obtained from the asymptotic slope of mean square distance, has negligible uncertainty in its slope and hence is not mentioned. The quantum mechanical calculations have no statistical error associated with them and can be obtained to any degree of accuracy. 

\subsection{Fluctuations in ACF and statistical error associated with Green Kubo approach}

\begin{figure}
    \centering
    \includegraphics[trim = 40 0 43 28,clip, scale=0.3]{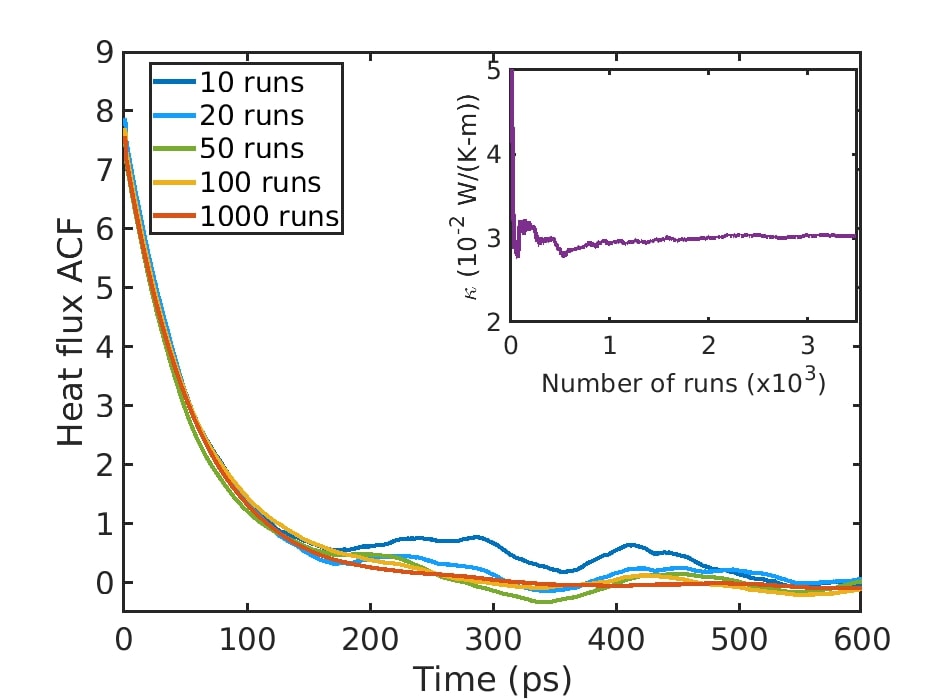}
    \caption{Heat flux autocorrelation function (ACF) calculated in a cumulative manner over several runs at 30 K and 0.10 MPa. The inset shows the thermal conductivity value calculated at the end of each run. }
    \label{fig:corr_avg}
\end{figure}

As mentioned in the previous sections, to minimize the error due to fluctuations in the ACF, it is calculated as a cumulative average of the previous runs. The random fluctuations in the ACFs are white noises which cancel out when averaged over a large data sample. Here a single run refers to the simulation duration over which one ACF is calculated. The duration of this changes for each simulation depending on the temperature and pressure conditions as the time the ACF takes to decay varies with these parameters. Fig. \ref{fig:corr_avg} shows how the heat flux ACF converges to an exponentially decaying function when cumulatively averaged over several runs. The inset in Fig. \ref{fig:corr_avg} shows the thermal conductivity calculated at the end of each run using direct integration of the cumulatively averaged ACF. 
\begin{figure}
    \centering
    \includegraphics[trim = 27 0 60 40,clip, scale=0.28]{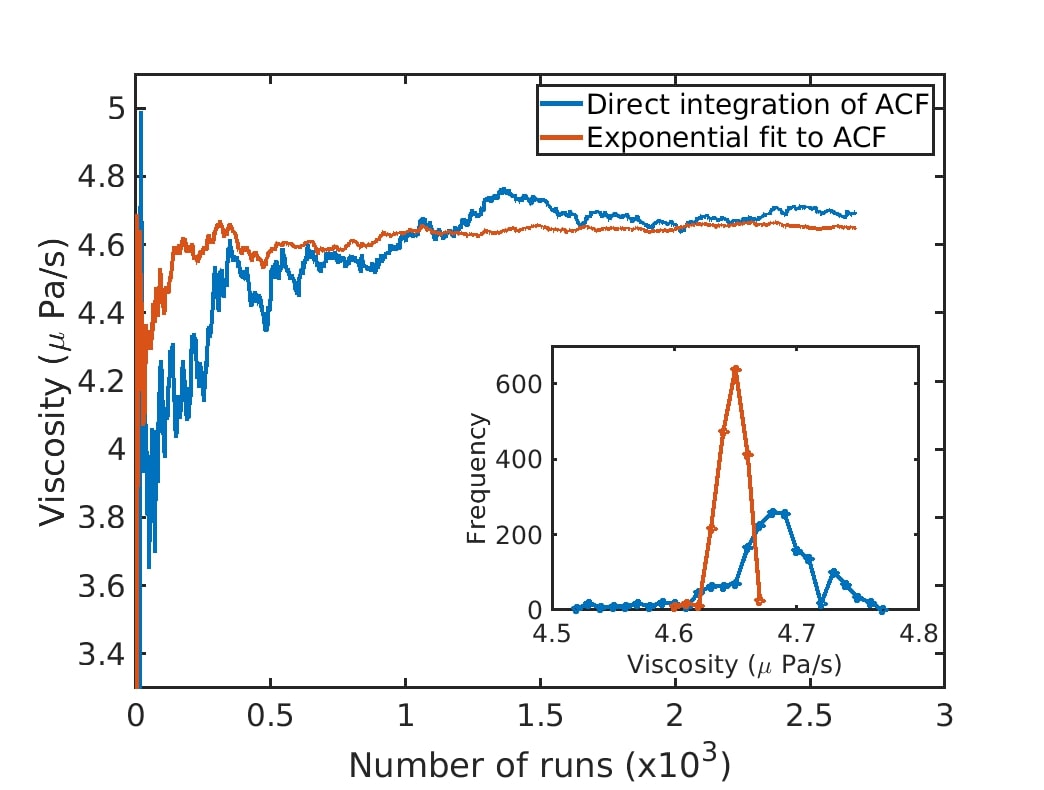}
    \caption{Viscosity calculated using direct integration of the autocorrelation and exponentially fitting autocorrelation function (ACF) at 30 K and 0.21 MPa. The inset shows the frequency distribution curve of values of viscosity calculated after 1500 runs. }
    \label{fig:exp_fit}
\end{figure}
\begin{center}
   \begin{figure}
         \centering
         \includegraphics[trim= 0 0 0 0, clip,scale=0.31]{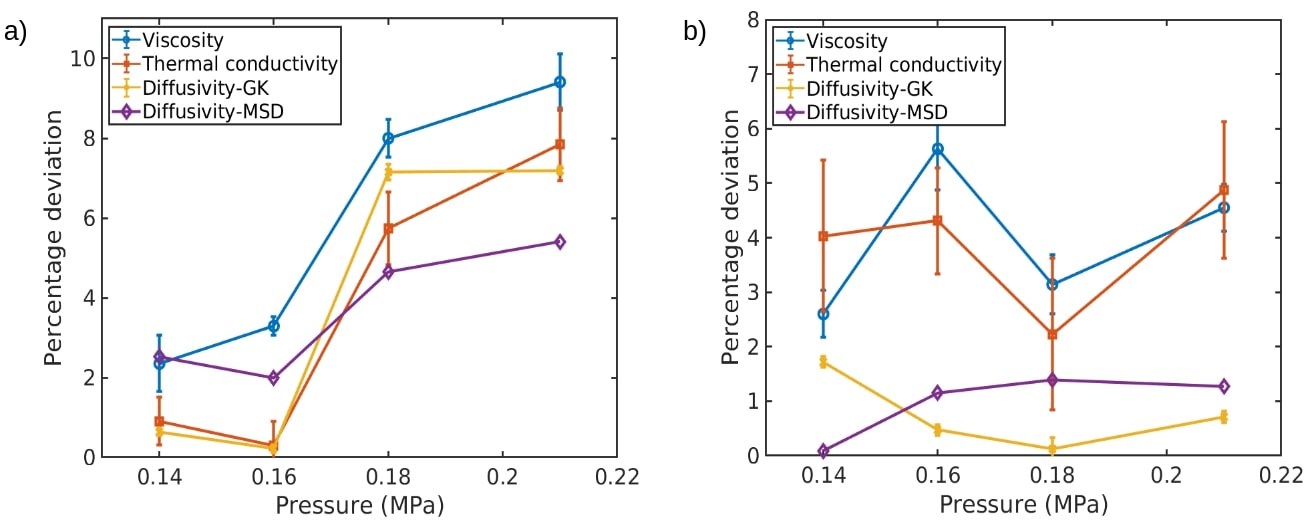}
         \caption{Percentage deviation in the calculated transport properties, from the quantum mechanically calculated values, for different pressures at, (a) 30 K and (b) 100 K.}
         \label{fig:qm_err}
\end{figure} 
\end{center}
In principle, an average over infinite data sets is required to remove white noise completely from the system. Hence to minimise any residual fluctuations in the ACF and errors due to cut-offs in the ACF integration, cumulatively averaged ACF is fit to an exponentially decaying function ($a\ e^{-t/\tau}$) and the integral over ACF is analytically given by $a\tau$. Fig. \ref{fig:exp_fit}, shows the viscosity calculated at 30 K and 0.21 MPa by direct integration (red line) of the cumulatively averaged ACF and by analytical integration of exponentially fit cumulatively averaged ACF. It can be observed that a better convergence is obtained in the calculated viscosity values. In the inset of Fig. \ref{fig:exp_fit}, the number of times a particular viscosity value appears when the simulation is run several times ($\sim 1500$ runs) is given in the form of a frequency distribution plot. The blue curve shows the frequency distribution for the values by direct integration (trapezoidal method) of ACF and the red curve shows the distribution for the values obtained by integration of the exponential fit. The width of the frequency distribution plot obtained for the exponentially fit ACF is lower than that obtained by direct integration. This shows a reduction in the standard deviation (statistical error) of the obtained viscosity values. It can also be observed that the two curves peak at different values. This shows that the exponential fit also accounts for the data lost from the tail of the ACF if an arbitrary cutoff is used.

\subsection{Uncertainty due to quantum effects}
\begin{figure}
         \centering
         \includegraphics[trim= 10 0 35 25, clip,scale=0.32]{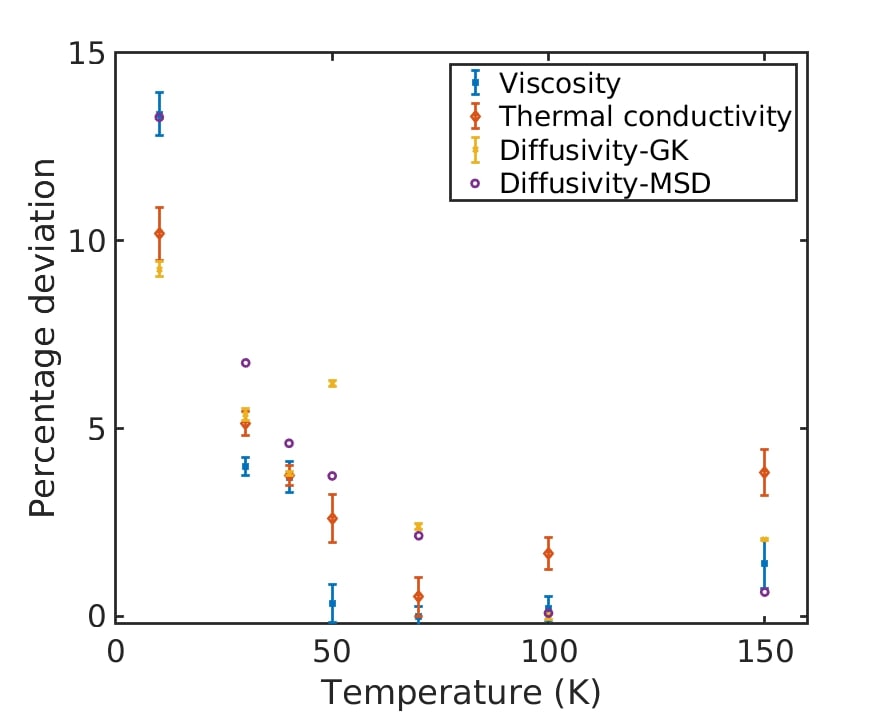}
         \caption{Percentage deviation in the calculated transport properties, from the quantum mechanically calculated values, at different temperatures at 0.10 MPa pressure.}
         \label{fig:temp}
\end{figure}

Fig. \ref{fig:qm_err} (a) and (b) show the percentage deviation in the transport coefficients, obtained through (I) MDS and (II) from the quantum mechanically calculated values at different temperatures and pressures. Fig \ref{fig:temp} shows how the deviation in the transport coefficients increases with a decrease in temperature. Below 40 K, at 0.10 MPa, the quantum mechanical effects are considerable, causing deviation in the classically simulated values. In Fig. \ref{fig:qm_err} (a), at a temperature of 30 K, the deviations increase with the increase in pressure. This can be attributed to the fact that with an increase in pressure, the mean interparticle distance decreases and hence approaches the $\lambda_{th}$ at higher temperatures. In Fig. \ref{fig:qm_err} (b), the deviations in the transport coefficients remain roughly the same with pressure. This shows that the pressure range considered is not high enough to make the quantum effects prominent at 100 K. It is also evident that the uncertainties due to quantum effects are present independent of the calculation procedure, GK or MSD, used within the classical MDS.
\section{Discussion}
In this article, the transport properties such as viscosity, diffusion coefficient, and thermal conductivity of helium gas are calculated in the temperature range of 10 K to 150 K and the pressure range of 0.10 MPa to 0.21 MPa. Temperatures below the critical temperature (5.2 K), where the effects due to quantum statistics are considerable, and pressures above the critical pressure (0.22 MPa), where the $^4$He is in the supercritical fluid state are not considered in this study.

The most prominent sources of uncertainty in the transport coefficient calculations at cryogenic temperatures using the equilibrium MDS are the quantum effects and the statistical errors associated with the GK approach. From the simulations performed, it is observed that the uncertainties due to quantum effects increase with a decrease in temperature and an increase in pressure due to the mean interparticle distance approaching the thermal de Broglie wavelength. Due to fluctuations in the temperature, pressure and density values during the simulation, there is an uncertainty in the transport coefficient values computed. However, this uncertainty is less than 0.5\% and can be neglected in comparison to the statistical fluctuations in the ACF or the uncertainty due to the quantum effects.

The uncertainties from the GK approach can be minimised to some extent using different integration methods for the ACF but the uncertainties due to the quantum effects can't be removed as long as one confines themselves to the classical MDS. One can perform quantum MDS \cite{QMsim1,QMsim2}, which includes the quantum effects, but quantum simulations are orders of magnitude more computationally intensive than classical simulations.  Another way around this problem is to use existing quantum correction procedures for classical MDS, but they are limited in their applicability. The well-known Feynman-Hibbs (FH) variational approach, which accounts for the positional uncertainty of quantum particles \cite{feynmanhibbs,mod_dif}, has been used successfully for many studies such as the adsorption of hydrogen and deuterium \cite{fh_hd}, quantum effects in light and heavy water \cite{fh_water}, equation of state for solid neon \cite{fh_fpe} and transport properties of \textsuperscript 4He at 80 K, neon at 45 K and methane at 110 K \cite{methane}. This approach has its limitations. Although it works well in moderate to high temperature regimes (> 80 K, for the case of $^4$He), it breaks down in the temperature regime where the quantum diffraction effects \cite{methane}, explained in Sec. II B, are high (T < 40 K, for $^4$He).  This is attributed to the FH method's inability to correctly yield the kinetic energy of the particles in the system due to the increased delocalisation at lower temperatures \cite{fh_rev,Erich}. A worthwhile future research direction would be exploring different correction procedures to include quantum effects in the classical simulation methods at cryogenic temperatures.
 
\section{Acknowledgement}
We gratefully acknowledge the financial support from the funding agency Department of Science and
Technology (DST-SERB), Government of India, Grant No. CRG/2022/000112. The authors sincerely thank the High-Performance Computing facility (HPC) at IIT Kanpur, because of which this work has been made possible.
\section{Author Declaration}
\subsection{Conflict of Interest}
The authors declare no conflict of interest.
\subsection{Data Availability}
The data that support the findings of this study are available from the corresponding author upon reasonable request. 

\renewcommand{\arraystretch}{1.5} % Default value: 1
\begin{table}[H]
\caption{\centering Viscosity ($\nu$) ($\mu Pa/ s$) data obtained using quantum calculations (QM), and molecular dynamics simulation (MDS) with the statistical error associated with the Green Kubo approach.}
\centering
\begin{tabularx}{1\textwidth} {|
>{\centering\arraybackslash}X   | >{\centering\arraybackslash}X   ||  >{\centering\arraybackslash}X   | >{\centering\arraybackslash}X   ||}
\hline
\multirow{2}{6em}{\centering Temperature (K)} & \multirow{2}{6em}{\centering Pressure (MPa)} & \multicolumn{2}{c||}{Viscosity ($\nu$) }\\
\cline{3-4}
&& QM  & MDS (GK)\\
\hline
10&\multirow{7}{2em}{0.10} & \num{2.02}& \num{1.55}$\pm$0.01\\
30 &&\num{4.25} & \num{4.08}$\pm$0.01\\
40 && \num{5.12} & \num{4.93 }$\pm$0.02\\ 
50 &&\num{5.91}& \num{5.93 }$\pm$0.03 \\ 
70 && \num{7.34} & \num{7.34}$\pm$0.02\\ 
100 && \num{9.23}& \num{9.25}$\pm$0.03\\ 
150 && \num{12.1}& \num{11.9}$\pm$0.1\\ 
\hline
\multirow{4}{1em}{30} 
&0.14&\multirow{4}{1.75em}{\num{4.25 }}&\num{4.35 }$\pm$0.03\\
&0.16&& \num{4.39 }$\pm$0.01\\
& 0.18&& \num{4.59 }$\pm$0.02\\ 
& 0.21&& \num{4.65 }$\pm$0.03\\ 
\hline
\multirow{4}{1.5em}{100} 
&0.14&\multirow{4}{1.75em}{\num{9.23}}&\num{9.47 }$\pm$0.04\\
& 0.16 && \num{9.75 }$\pm$0.07\\
& 0.18 && \num{9.52 }$\pm$0.05\\ 
& 0.21 && \num{9.65 }$\pm$0.04\\ 
\hline
\end{tabularx}
\label{tab:V_data}
\end{table}
%%%%%%%%%%%%%%%%%%%%%%%%%%%%%%%%%%%%%%%%%%%%%%%%%%
%\newpage
%\section{Transport Coefficients Data}
\renewcommand{\arraystretch}{1.5} % Default value: 1
\begin{table}[H]
\caption{\centering Thermal conductivity ($\kappa$) ($10^{-2}W/(K-m)$) data obtained using quantum calculations (QM), and molecular dynamics simulation (MDS) with the statistical error associated with the Green Kubo approach.}
\centering
\begin{tabularx}{1\textwidth} {|
>{\centering\arraybackslash}X   | >{\centering\arraybackslash}X   ||  >{\centering\arraybackslash}X   | >{\centering\arraybackslash}X   ||}
\hline
\multirow{2}{6em}{\centering Temperature (K)} & \multirow{2}{6em}{\centering Pressure (MPa)} & \multicolumn{2}{c||}{Thermal conductivity ($\kappa$)}\\
\cline{3-4}
&& QM  & MDS (GK)\\
\hline
10&\multirow{7}{2em}{0.10} & \num{1.57}&\num{1.02}$\pm$0.01\\
30 & &\num{3.31}&\num{3.14}$\pm$0.01\\
40 && \num{3.99 }& \num{3.84 }$\pm$0.01\\ 
50 && \num{4.61}& \num{4.73}$\pm$0.03\\ 
70 && \num{5.72}& \num{5.69 }$\pm$0.03\\ 
100 &&\num{7.19}& \num{7.07}$\pm$0.03\\ 
150 && \num{9.40}& \num{9.76}$\pm$0.06\\ 
\hline
\multirow{4}{1em}{30} 
&0.14&\multirow{4}{1.75em}{\num{3.31 }}&\num{3.34 }$\pm$0.02\\
&0.16&& \num{3.32}$\pm$0.02\\
& 0.18&& \num{3.50 }$\pm$0.03\\ 
& 0.21&& \num{3.57 }$\pm$0.02\\ 
\hline
\multirow{4}{1.5em}{100} 
&0.14&\multirow{4}{1.75em}{\num{7.19}}&\num{7.5 }$\pm$0.1\\
& 0.16 && \num{7.50 }$\pm$0.07\\
& 0.18 && \num{7.4 }$\pm$0.1\\ 
& 0.21 && \num{7.54 }$\pm$0.09\\ 
\hline
\end{tabularx}
\label{tab:K_data}
\end{table}
%\newpage
%%%%%%%%%%%%%%%%%%%%%%%%%%%%%%%%%%%%%%%%%%%%%%%%%%%%%%%%%%%%%%%%%%%%%%%%
\renewcommand{\arraystretch}{1.5} % Default value: 1
\begin{table}[H]
\caption{\centering Diffusion coefficients ($\mathcal{D}$) ($10^{-6}m^2/s$) data obtained using numerical quantum calculations (QM), and molecular dynamics simulation (MDS) with the statistical error associated with the Green Kubo approach.}
\centering
\begin{tabularx}{1\textwidth} {|>{\centering\arraybackslash}X   | >{\centering\arraybackslash}X||  >{\centering\arraybackslash}X|>{\centering\arraybackslash}X| >{\centering\arraybackslash}X   ||}
\hline
\multirow{2}{6em}{\centering Temperature (K)} & \multirow{2}{6em}{\centering Pressure (MPa)} & \multicolumn{3}{c||}{Diffusivity ($\mathcal{D}$)}\\
\cline{3-5}
&& QM  & MSD & MDS (GK)\\
\hline
10&\multirow{7}{2em}{0.10} &\num{0.55}&\num{0.47}&\num{0.50}$\pm$0.001 \\
30 & &\num{3.48}& \num{3.25}&\num{3.29}$\pm$0.005\\
40 && \num{5.61}& \num{5.35}&\num{5.40}$\pm$0.003\\
50 && \num{8.11}& \num{7.81} &\num{7.61}$\pm$0.006\\ 
70 && \num{14.1}& \num{13.8}&\num{13.8}$\pm$0.01\\ 
100 && \num{25.4 }& \num{25.4}&\num{25.4}$\pm$0.02\\ 
150 && \num{49.6}& \num{49.3}&\num{50.6}$\pm$0.1\\ 
\hline
\multirow{4}{1em}{30} 
&0.14&\num{2.52}&\num{2.44}&\num{2.50}$\pm$0.002\\
&0.16& \num{2.20}&\num{2.16}&\num{2.21}$\pm$0.002\\
& 0.18& \num{1.96} &\num{2.02}&\num{2.10}$\pm$0.004\\ 
& 0.21& \num{1.68} &\num{1.77}&\num{1.80}$\pm$0.001\\ 
\hline
\multirow{4}{1.5em}{100} 
&0.14&\num{18.4 }&\num{18.4}&\num{18.1}$\pm$0.02\\
& 0.16 & \num{15.8 }&\num{16.1}&\num{16.0}$\pm$0.02\\
& 0.18 & \num{14.1 } &\num{14.3}&\num{14.3}$\pm$0.02\\ 
& 0.21 & \num{12.1 } &\num{12.3}&\num{12.3}$\pm$0.01\\ 
\hline
\end{tabularx}
\label{tab:D_data}
\end{table}
%%%%%%%%%%%%%%%%%%%%%%%%%%%%%%%%%%%%%%%%%%%%%%%%%%%%%%%%%%%%%%%%%%%%%%%%%%%%%%%%
\appendix
\section{Additional details on the simulations performed}

The details of the number of simulation runs performed and thereby, the number of data points collected for the transport properties at each state point, using the GK method, are provided in Table \ref{tab:data_sim}.

\renewcommand{\arraystretch}{1.5} % Default value: 1
\begin{table}[H]
\caption{\centering Number of data points collected using the GK method for each of the transport properties at different state points.}
\centering
\begin{tabularx}{1\textwidth} {|>{\centering\arraybackslash}X   | >{\centering\arraybackslash}X||  >{\centering\arraybackslash}X|>{\centering\arraybackslash}X|>{\centering\arraybackslash}X||}
\hline
\multirow{2}{6em}{\centering Temperature (K)} & \multirow{2}{6em}{\centering Pressure (MPa)} & \multicolumn{3}{c||}{Number of data points}\\
\cline{3-5}
&& Viscosity ($\nu$)  & Thermal conductivity ($\kappa$)&Diffusivity ($\mathcal D$) \\
\hline
10&\multirow{7}{2em}{0.10} &\num{3309}&\num{1379}& \num{1017}\\
30 & &\num{8400}& \num{3500}& \num{1690}\\
40 && \num{8400}& \num{3500}& \num{3333}\\
50 && \num{8400}& \num{1750}& \num{3333} \\ 
70 && \num{8000}& \num{4444}& \num{3333}\\ 
100 && \num{6251}& \num{3333}& \num{3333}\\ 
150 && \num{8000}& \num{2500}& \num{3333}\\ 
\hline
\multirow{4}{1em}{30} 
&0.14&\num{3974}&\num{1655}& \num{3195}\\
&0.16& \num{3432}&\num{1430}& \num{2670}\\
& 0.18& \num{2661} &\num{1108}& \num{2060}\\ 
& 0.21& \num{2670} &\num{1112}& \num{2218}\\ 
\hline
\multirow{4}{1.5em}{100} 
&0.14&\num{1750 }&\num{4200}& \num{3333}\\
& 0.16 & \num{1750}&\num{4200}& \num{3333}\\
& 0.18 & \num{1750} &\num{4200}& \num{3333}\\ 
& 0.21 & \num{1750} &\num{4200}& \num{3333}\\ 
\hline
\end{tabularx}
\label{tab:data_sim}
\end{table}
\bibliographystyle{elsarticle-num} 
 \bibliography{cas-refs}
\end{document}